\documentclass[%
preprint,
 amsmath,amssymb,
 aps,
]{revtex4-2}

\usepackage{graphicx}
\usepackage{dcolumn}
\usepackage{bm}

\usepackage{pifont}
\usepackage{xcolor}
\newcommand{\cmark}{\textcolor{green!80!black}{\ding{51}}}
\newcommand{\xmark}{\textcolor{red}{\ding{55}}}
\usepackage{placeins}
\usepackage{nameref}
\usepackage{siunitx}
\usepackage{cleveref}
\renewcommand{\vec}{\boldsymbol}
\begin{document}


\title{Patterns of active dipolar particles in external magnetic fields}

\author{Vitali Telezki }
\author{Stefan Klumpp }%
 \email{stefan.klumpp@phys.uni-goettingen.de}
\affiliation{%
 University of G\"ottingen, Institute for the Dynamics of Complex Systems,  Friedrich-Hund-Platz 1, 37077 G\"ottingen, Germany
}%



\date{\today}

\begin{abstract}
Active particles with a (magnetic) dipole moment are of interest for steering self-propelled motion, but also result in novel collective effects due to their dipole-dipole interaction. Here systems of active dipolar particles are studied with Brownian dynamics simulations to systematically characterize the different patterns they form, specifically in the presence of an external (magnetic) field. The combination of three types of order  - clustering, orientational alignment and chain formation - is used to classify the patterns observed in these systems. In the presence of an external field, oriented chains and bands are found to be dominant.  These structures show some similarities with columnar cluster seen in (passive) ferrofluids and display columnar spacing and number of lanes per cluster that both decrease with increasing field strength.
\end{abstract}

\maketitle
 \newpage


\section{Introduction}
Systems of self-propelled particles as a prime example of active matter have been studied extensively for their complex collective behavior \cite{Vicsek2012,Romanczuk2012,Bechinger2016}, both from the purely theoretical point of view where these systems provide examples for systems intrinsically out of equilibrium \cite{Cates2015a} and from an applied point of view where self-propelled  particles are model systems for microrobots with biomedical and environmental applications \cite{Bente2018}. From the applications point of view, the remote control of active particles is an important feature. An important method for remote control is through magnetic fields, provided that the active particles are equipped with a magnetic dipole moment \cite{Bente2018,Klumpp2019}. Active particles with a magnetic dipole moment can be found in nature, in the form of magnetotactic bacteria \cite{Klumpp2019,Bazylinski2004}, but can also be made fully synthetically, e.g. active magnetic colloids \cite{baraban2012catalytic,alsaadawi2021control} and various forms of magnetic propellers and microrobots \cite{ghosh2009controlled,vach2015fast}, and as biohybrid systems by magnetically functionalizing motile microorganisms \cite{carlsen2014magnetic,magdanz2013development}. 

While a large body of work has explored the magnetic steering of active particles as well as magnetically driven propulsion mechanisms \cite{Klumpp2019}, the effects of the magnetic interactions, both among the active particles and with an external magnetic field, have received considerably less attention.  
A few studies have explored the standard two-dimensional active Brownian particle model with additional magnetic interactions based on magnetic moments in the same plane as the active motion. Liao et al.~\cite{Liao2020} have determined the states of active particles with dipolar interactions in the absence of an external field and shown various patterns, in particular aggregation into chains. Similarly, our own earlier work has addressed patterns in small systems with only a few particles in confinement and studied the emergence of chain and rings \cite{Telezki2020}. These and additional structures were also found computationally as well as experimentally using centimeter-sized toy robots in a recent study with small particle numbers and harmonic confinement \cite{obreque2024dynamics}.  Moreover, motility-induced phase separation was shown to be suppressed by the magnetic interactions \cite{Liao2020,sese2022impact}. Mixtures, in which dipolar particles are not active themselves, but interact with active particles, have also been studied and also show aggregation into chains, but also large-scale clustering of the particles \cite{Maloney2020,Maloney2020a}. Finally, magnetic interactions were shown to result in the clustering of magnetotactic bacteria in flow through narrow channels \cite{waisbord2016destabilization,meng2018clustering}.

In this study,  we extend the previous work \cite{Liao2020,Telezki2020} to study the patterns that emerge in systems of dipolar active particles in the presence of an external field. We use systematic computer simulations to study the ordered states in such system as a function of the density, field strength, dipole moment and activity. We distinguish up to eight different states based on three order parameters introduced by Liao et al.~\cite{Liao2020} that describe the competition between chaining, clustering and orientation. As a reference, we start with the case without the external field and study how a magnetic field modulates these states. We show that under strong fields, bands of oriented particles are formed that constitute active particle equivalents of columnar clusters know from passive ferrofluids \cite{Liu1995,Flores1999,Ivey2001}. 

\section{Methods}
\label{ch:methods}

\subsection{Model}
\begin{figure}
    \centering
    \includegraphics{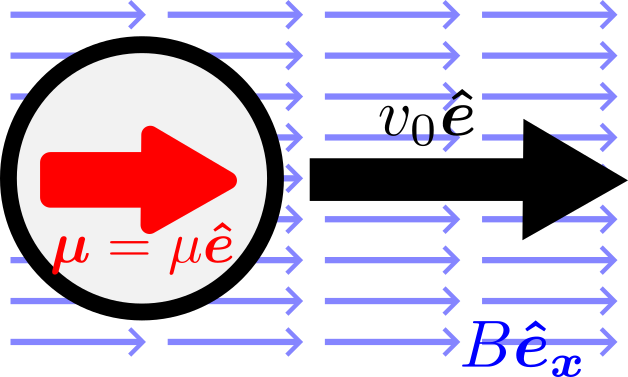}
    \caption{Schematic illustrating single active particle with orientation $\hat{\bm{e}} $ in external homogeneous magnetic field $B\hat{\bm{e}}_x$ (blue). Particle has magnetic strength $\mu$ and self-propulsion speed $v_0$ (back). }
    \label{fig:schematicSwimmer}
\end{figure}

We consider spherical active Brownian particles in two dimensions that have a permanent dipole moment 
$\vec \mu_i = \mu \hat{\vec e}_i$ with the magnetic strength $\mu$ at the particle center. Here and in the following, the superscript 'hat' denotes a unit vector. The magnetic moment is aligned
with the orientation of the particle $\hat{\vec e}_i$, which defines the direction of self-propulsion with speed $v_0$ (\cref{fig:schematicSwimmer}).
The equations of motion for a particle $i$ with position $\vec r_i$ and orientation $\hat{\vec e}_i$ are given by
\begin{align}
    \dot{ \vec r_i} &= v_0 \hat{\vec e}_i + \frac{1}{\gamma_\mathrm{T}}\vec F_i + \sqrt{2D_{\mathrm{T}}}\vec\xi_i^{\mathrm{T}} \label{eq:motionPos} \\
    \dot{\hat{\vec e}}_i &= \frac{1}{\gamma_\mathrm{R}}\vec{\tau}_i \times \hat{\vec{e}}_i +\sqrt{2D_{\mathrm{R}}} \vec\xi_i^{\mathrm{R}} \times \hat{\vec e}_i, \label{eq:motionOrient} 
\end{align}
where $\gamma_\mathrm{T}$ and $\gamma_\mathrm{R}$ are the translational and rotational drag coefficients, $\vec\xi_i^{\mathrm{T}}$ is the translational stochastic force, and 
$\vec\xi_i^{\mathrm{R}}$ is the rotational stochastic torque.
Both are described by Gaussian white noise with zero mean
\begin{align*}\left<\xi^{\alpha}_i\!\left(t\right)\xi^{\beta}_j\!\left(t^\prime\right) 
\right>&=\delta_{i,j}\delta_{\alpha,\beta}\delta\!\left(t-t^\prime\right),
\end{align*}
where the upper index ($\alpha, \beta$) runs over the $x$- and $y$-components of the translational noise as well as the rotational noise ($z$ component), and the lower index ($i,j$) runs over the particles. 
$D_\mathrm{T}$ and $D_\mathrm{R}$ are the corresponding diffusion coefficients. $\vec{F}_i$ and $\vec{\tau}_i$
describe forces and torques acting on particle $i$.
The forces $\vec{F}_i$ acting on particle $i$ are given by the sum of the particle-particle interactions 
\begin{align*}
    \vec{F}_i = \sum_{j\neq i} \vec{F}^{\mathrm{dd}}_{ij}  + \sum_{j\neq i}\vec{F}^{\mathrm{WCA}}_{ij}
\end{align*}
The pairwise forces between two particles $i$, $j$ separated by the distance 
$r_{ij} = \left|\vec{r}_{ij}\right| = \left| \vec{r}_i - \vec{r}_j \right|$ 
consist of dipolar forces and excluded volume interactions. 
The dipolar forces are given by~\cite{Yung1998}
\begin{equation*}
\begin{aligned}
    \vec{F}^{\mathrm{dd}}_{ij}     =& \frac{3\mu_0 \mu^2}{4\pi r_{ij}^4}
    \left[
    \hat{\vec r}_{ij} \left( \hat{\vec e}_i \cdot \hat{\vec e}_j \right) 
    + \hat{\vec e}_i \left( \hat{\vec r}_{ij} \cdot \hat{\vec e}_j \right)
    + \hat{\vec e}_j \left( \hat{\vec r}_{ij} \cdot \hat{\vec e}_i \right) \right. \\
    & \left.- 5 \hat{\vec r}_{ij} \left( \hat{\vec r}_{ij} \cdot \hat{\vec e}_i \right)
    \left( \hat{\vec r}_{ij} \cdot \hat{\vec e}_j \right) \vphantom{\hat{vec r}_{ij}}
    \right].
\end{aligned}
\end{equation*}
The excluded volume interactions are modeled by the Weeks-Chandler-Anderson (WCA)~\cite{Chandler1983}
pair potential with the interaction energy $\epsilon$ and the particle diameter $\sigma$
\begin{align*}
    \vec{F}^{\mathrm{WCA}}_{ij} = 
    \begin{cases}
    48 \epsilon \frac{\vec r_{ij}}{r_{ij}^2}\left[
    \left( \frac{\sigma}{r_{ij}} \right)^{12} - \frac{1}{2} \left( \frac{\sigma}{r_{ij}} \right)^{6} \right]  &
    \text{if $r_{ij}<2^{1/6}\, \sigma $} \\
     0 &
     \text{if $r_{ij} \geq 2^{1/6}\,\sigma$}.
\end{cases}
\end{align*}

The torques $\vec{\tau}_i$ acting on the particle $i$ result from 
the pairwise dipolar interactions \cite{Yung1999} and the interaction with the external homogeneous magnetic field ${\vec B}=B\hat{\vec e}_x$ (stationary frame of reference), 
\begin{align*}
    \vec{\tau}_i &= \sum_{j \neq i} \vec{\tau}^{\mathrm{dd}}_{ij} + \mu \hat{\vec e}_i \times {\vec B} \\
    &= \sum_{j \neq i} \frac{\mu_0\mu^2}{4\pi r_{ij}^3}
    \left[
    3\left(\hat{\vec e}_i \times \hat{\vec r}_{ij} \right)
    \left(\hat{\vec e}_j \cdot \hat{\vec r}_{ij} \right)  -
    \left(\hat{\vec e}_i \times \hat{\vec e}_j \right)
    \right] + \mu \hat{\vec e}_i \times {\vec B}.
\end{align*}

\subsection{Simulation method}
The dynamics of active dipolar particles were simulated using the 
simulation toolkit HOOMD-blue, version 2.6 \cite{Anderson2008,Anderson2020}. The active
self-propulsion of the particles and their alignment in an external magnetic field were added through modules developed in-house. The project data and its parameter space was managed within a signac framework \cite{Adorf2018,Ramasubramani2018}. 

All simulations were carried out using $N=1156$ particles placed in a quadratic simulation box with side length $L_x =L_y =L$ with periodic boundary conditions. This relatively small system size was chosen to keep the simulation time manageable while systematically scanning a rather large parameter space. The same system size was used in previous work without an external field, where it was also shown that cluster sizes saturated for $N>1000$ \cite{Liao2020}. To
avoid interactions between periodic images of the particles, a cut-off distance of the order of half the system size, $r_{\mathrm{cut}}=L/2-\sigma$, was introduced for dipolar 
interactions. To recover correct rotational diffusion dynamics for spheres, we set $\gamma_R = 3\gamma_T$ within HOOMD-blue. To study the effect of density, we varied the lateral size $L$ of the simulation box. We chose the following three density values defined by the mean area fraction $\Phi = N\pi\sigma^2/\left(4L^2\right)$: 
One well below ($\Phi = \num{0.13},\,L=\num{85}$), one close to ($\Phi = \num{0.23},\, L=\num{63}$) and one above ($\Phi = \num{0.57},\, L=\num{40}$) the critical density where MIPS is expected in systems of active Brownian particles without dipolar interactions. 
To investigate the effect of a constant homogeneous external field on the collective dynamics of 
dipolar active particles, we systematically explored different combinations of the strength $B$ of the external field and the magnetic moment of the particles. 

As in our previous work \cite{Telezki2020}, we used the following dimensionless units: 
distances are measured in units of the particle size $\sigma$ and time in units of $\sigma^2/D_0^T$, the time in which a particle diffuses over a distance of its own size. The strength of the WCA interaction $\epsilon$ is used as the unit of energy, thus temperature $T$ has units of $\epsilon/k_B$ (here, $T=1$). All magnetic quantities are expressed relative to a reference field strength $B_0=1\times 10^{-5}$~T, which is of the order of the magnetic field of the Earth. Thus, the strength of the magnetic field has units of $B_0$ and magnetic moments are expressed units of $\epsilon/B_0$. Note that with these units, the self-propulsion velocity $v_0$ is identical to the Peclet number $v_0\sigma/D_0^T$. 
Numerical integration was performed for
$N_t=10^6$ time steps with an integration time step width $\delta_t=2\times10^{-5}$. Simulations were evaluated 
from $t=18$ to $t=20$ with data taken every $\Delta t=0.1$.

\subsection{Order parameters}
\label{ch:globalOrder}

In large systems of active dipolar particles, complex structures such as branches, connected networks, bands or 
loops within rings and large scale cluster formation are expected to emerge 
\cite{Klapp2005,Klapp2005,Liao2020,Maloney2020,Maloney2020a,Holm2005}.
To characterize such patterns, we used the following three global order parameters introduced by Liao et al.~\cite{Liao2020}.
We used combinations of these three order parameters to identify and classify different state of the system and to construct state diagrams as described below. 

Similar to other active systems, we expect active dipolar particles to form large clusters. To quantify cluster formation,
we determined the fraction of all particles that are part of the largest cluster by counting the number of particles  $n^*_{\mathrm{c}}$ in that cluster relative to the total particle number of the system $N$,
\begin{equation}
    \phi_{\mathrm c} = \frac{n^*_{\mathrm{c}}}{N}.
\end{equation}
The fraction of the largest cluster is close to 0 when particles form small or no clusters and reaches 1 when the whole system forms one giant cluster. To identify clusters and, specifically, the largest cluster, we performed a density based cluster analysis (DBSCAN \cite{Pedregosa2011}) with the value of the cut-off distance $\delta_c$ based on the nearest neighbor distance $g_1(r)$ for each corresponding density. 

Orientational order due to dipolar interactions of alignment with the external field is characterized by the global polarization, which is given by the average over the orientations of all particles in the system
\begin{equation}
     \phi_{\mathrm e}  = \frac{1}{N}\left|\sum_i \hat{\vec e}_i\right|.
\end{equation}

Finally, to assess the degree of chain formation in large systems, we applied the following chain criteria, following Liao et al.~\cite{Liao2020}. Two particles are considered to be bonded if they fulfill the following three criteria: 
First, their distance is smaller than the cut-off distance, $r_{ij} < \delta_c$. Second, they are 
aligned in parallel $\hat{\vec e}_i \cdot \hat{\vec e}_j > 0$, and third their orientation is in line with their 
connecting distance vector $\left(
\hat{\vec e}_i\times \hat{\vec r}_{ij}\right) \cdot \left(\hat{\vec e}_j \times \hat{\vec r}_{ij}\right) >0$. 
The fraction of particles that fulfill all three criteria give the degree of polymerization
\begin{align} 
\phi_{\mathrm p} & =\frac{ n_{\mathrm{chain}} } {N}.
\end{align}

\section{Results}

In this study, we investigated how a homogeneous external magnetic field affects structure formation and collective behavior of 
(magnetic) dipolar active particles. To this end,  we performed extensive Brownian dynamics simulations of active dipolar particles 
in a homogeneous external magnetic field while systematically varying the field strength $B$, the active velocity $v_0$, and the magnetic moment $\mu$ (all in dimensionless units, see Methods). We performed our simulations for three representative densities $\Phi=$\numlist{0.13;0.23;0.57}. We classified the different states of the system with respect to their clustering, orientational order, and chain formation, summarizing the results in state diagrams. 

\subsection{No external field}
\begin{figure}[tbp]
    \centering
    \includegraphics[width=0.65\linewidth]{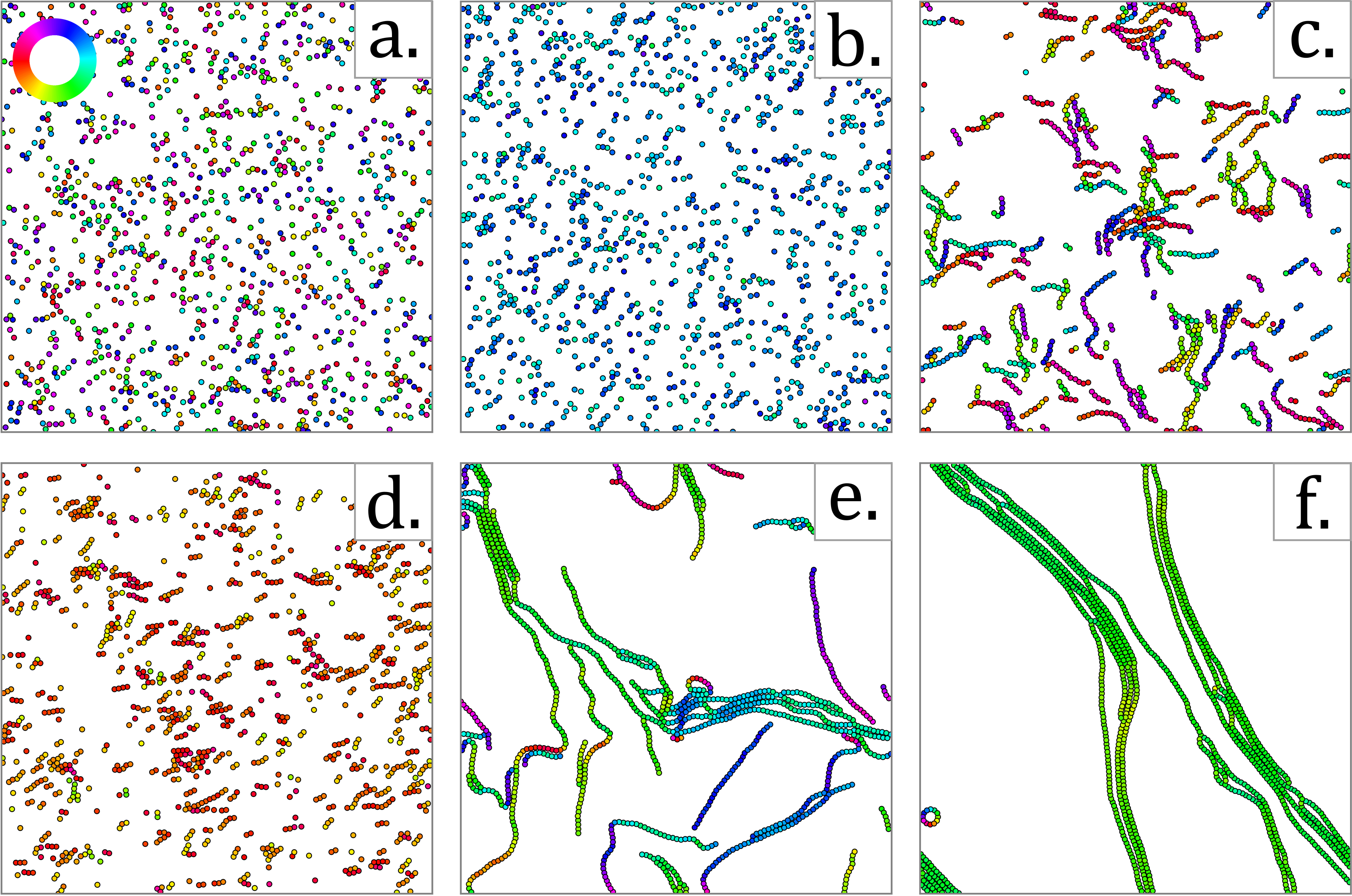}
    \caption{Snapshots showing examples for states introduced in \cref{tab:orderParamCombi} for systems at 
    low density ($\Phi=0.13$), without an external field. Snapshots were taken at time $t=20$. The following states are shown: a. disordered gas ($\mu=0$, $v_0=23$), b. oriented gas ($\mu=2.6$, $v_0=500$), c. gas of chains ($\mu=1.6$, $v_0=23$),  d. oriented chains ($\mu=1.6$, $v_0=500$), e. network of chains ($\mu=2.6$, $v_0=23$), f. bands ($\mu=4$, $v_0=100$).  The color wheel (top left) indicates the orientation of the particles.}
    \label{fig:snapshotsStatesOverview}
\end{figure}

As a reference, we first simulated active dipolar particles without an external magnetic field 
($B=0$). Here, we briefly present our results, in particular how the observed states depend on the activity $v_0$ and the pairwise 
magnetic interactions via the magnetic moment $\mu$.
This system has previously been analyzed in detail by Liao et al.~\cite{Liao2020} and the following results are in agreement with their results. In simulations for different $v_0$ and $\mu$ and different densities $\Phi$, we observe different patterns. Representative snapshots are shown in  \cref{fig:snapshotsStatesOverview}. These snapshots reflect combinations of three different types of ordering: alignment of the magnetic moments, clustering of the particles, and chain formation. To quantify the order in these patterns, we make use of combinations of the three previously introduced global order parameters, 
the fraction of particles in the largest cluster $\left\langle\phi_c\right\rangle$, the global polarization
 $\left\langle\phi_e\right\rangle$, and the degree of polymerization $\left\langle\phi_p\right\rangle$ (see Methods). 
All three parameters are dimensionless, ranging from 0 to 1. For systematic classification we define a system as ordered when its order parameter exceeds 0.5, i.e., when a majority of particles adopt the corresponding ordering. This threshold provides a rigorous criterion for order. By combining the three binary (ordered/disordered) states, eight distinct state classifications arise, as summarized in \cref{tab:orderParamCombi}. We note that this classification is more detailed than the one used in previous work~\cite{Liao2020}, even though it is based on the same order parameters. It allows us to systematically distinguish different types of order, which will become crucial in the presence of an external field as studied below.

\begin{figure}
\includegraphics[width=\linewidth]{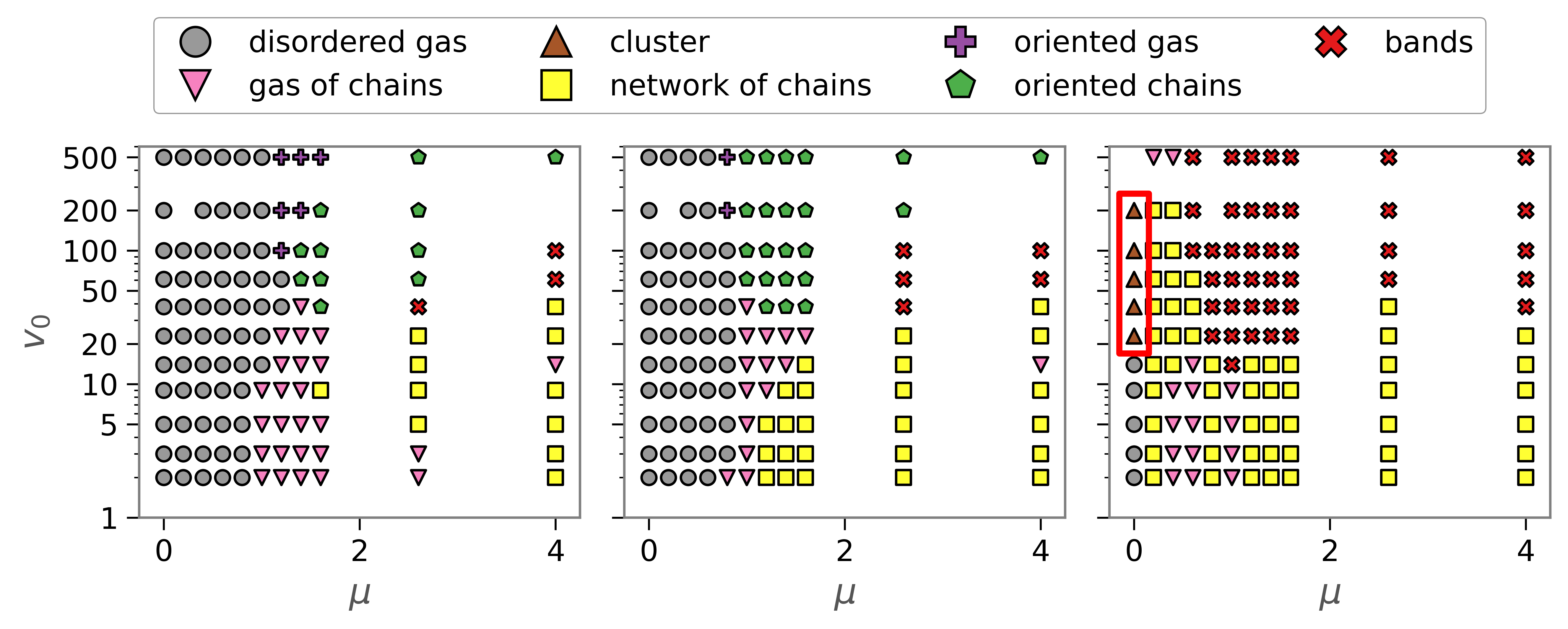}
\caption{Diagram of states in the absence of an external field for three densities $\Phi=$\numlist{0.13;0.23;0.57} (left to right). The states result from order parameter
criteria combinations according to \cref{tab:orderParamCombi}. The red box in the right diagram ($\Phi=0.57$) highlights cluster formation that  corresponds to MIPS for apolar ($\mu=0$) active particles.  }
\label{fig:diagramOfStatesNoB}
\end{figure}

\begin{table}[tb]
    \centering
    \begin{tabular}{c c c l}
    $\left\langle \phi_c\right\rangle>0.5$ & $\left\langle \phi_e\right\rangle>0.5$ & $\left\langle \phi_p\right\rangle>0.5$ & state\\[0.5ex]
    \hline 
      \cmark &  \xmark   &\xmark  & cluster\\
      \xmark &  \cmark   &\xmark  & oriented gas \\
      \xmark &  \xmark   &\cmark  & gas of chains \\
      \cmark &  \cmark   &\xmark  & oriented cluster\\
      \cmark &  \xmark   &\cmark  & network of chains\\
      \xmark &  \cmark   &\cmark  & oriented chains\\
      \cmark &  \cmark   &\cmark  & bands\\
      \xmark &  \xmark   &\xmark  & disordered gas\\
      \hline\hline
    \end{tabular}
    \caption{Order parameter combinations and their corresponding states. }
    \label{tab:orderParamCombi}
\end{table}

At the low and intermediate densities we simulated, we can distinguish six of these states that are shown in the snapshots in \cref{fig:snapshotsStatesOverview}: the disordered gas state with no order, the oriented gas with only orientational order, the gas of chains with only chain formation, oriented chains (chain formation and orientational order), a network of chains based on the combination of chain formation and clustering, and bands in the case where all three types of order are combined. 
 At high system density, well above the critical density for motility induced phase separation (MIPS, 
$\Phi_{\textnormal{MIPS}}\approx0.28$ \cite{Bruss2017}), we also see cluster formation in the apolar limit ($\mu = 0$) that corresponds to MIPS, as 
expected \cite{Liao2020}. The eighth possible state, oriented clusters, characterized by clustering and orientational order, but no chaining is not seen in the absence of an external field.

Based on the systematic classification, we summarized our results in the diagrams of states shown in  \cref{fig:diagramOfStatesNoB}.
For systems with densities below the critical density of MIPS ($\Phi=0.13$ and $\Phi=0.23$, \cref{fig:diagramOfStatesNoB} left and center) and 
weak magnetic interactions ($\mu < 1$) no classification criterion is fulfilled and the system is in a disordered
gas state (gray circle in \cref{fig:diagramOfStatesNoB}, snapshot in \cref{fig:snapshotsStatesOverview} a). Order (chain formation and/or global polarization) emerges for stronger magnetic interactions ($\mu\geq 1.0$).
At this point, contributions from dipolar interactions equal the energy
contributed by thermal noise ($\mu^2 = T = 1$) and induce chain formation, consistent with previous observations 
in small systems \cite{Telezki2020}. Different ordered states can
be realized  by varying the activity $v_0$ for fixed magnetic moment. For example, in the case of intermediate
magnetic interactions ($\mu = 1.4$) systems with low levels of activity tend to form a gas of
chains (pink triangle in \cref{fig:diagramOfStatesNoB}, snapshot  in \cref{fig:snapshotsStatesOverview} c). Increasing the activity adds global polarization to the chaining and, thus, results in oriented
chains (green pentagon in \cref{fig:diagramOfStatesNoB}, snapshot in  \cref{fig:snapshotsStatesOverview} d). For even higher levels of activity, the formation of
chains is suppressed while global polarization persists, so that an oriented gas is observed
(purple plus in \cref{fig:diagramOfStatesNoB}, snapshot in \cref{fig:snapshotsStatesOverview} b).

When the system density exceeds the critical density of MIPS ($\Phi=0.57$, \cref{fig:diagramOfStatesNoB} right), apolar ($\mu=0$)
and weakly magnetic particles ($\mu < 1$) also show ordered states like a network of chains (yellow square in \cref{fig:diagramOfStatesNoB}, snapshot in  
\cref{fig:snapshotsStatesOverview} e) or a gas of chains (pink triangle). In case of apolar particles, 
increasing the activity results in cluster formation (brown triangle) which has previously been identified as MIPS \cite{Liao2020} (and was confirmed here by a bimodal distributions of the local density). At high 
system density, network of chains (yellow square) and bands 
(red cross in \cref{fig:diagramOfStatesNoB}, snapshot in  \cref{fig:snapshotsStatesOverview} f) dominate the diagram of states.

\subsection{Influence of an external field}
With the definitions of order parameters and the state diagram for active dipolar particles without an external magnetic field in place as a reference point,
we investigated how a homogeneous external magnetic field affects those states. We first studied the lowest density simulated above ($\Phi=0.13$), results for higher densities will be reported below. 
Based on the expected average orientation of a reference system ($\mu=1$, $T=1$), we simulated three magnetic field 
strength, weak ($B=0.1$), intermediate ($B=1$) and strong ($B=14$), which we expect to represent different regimes.

\begin{figure}[!tpb]
    \centering
    \includegraphics[width=.6\columnwidth]{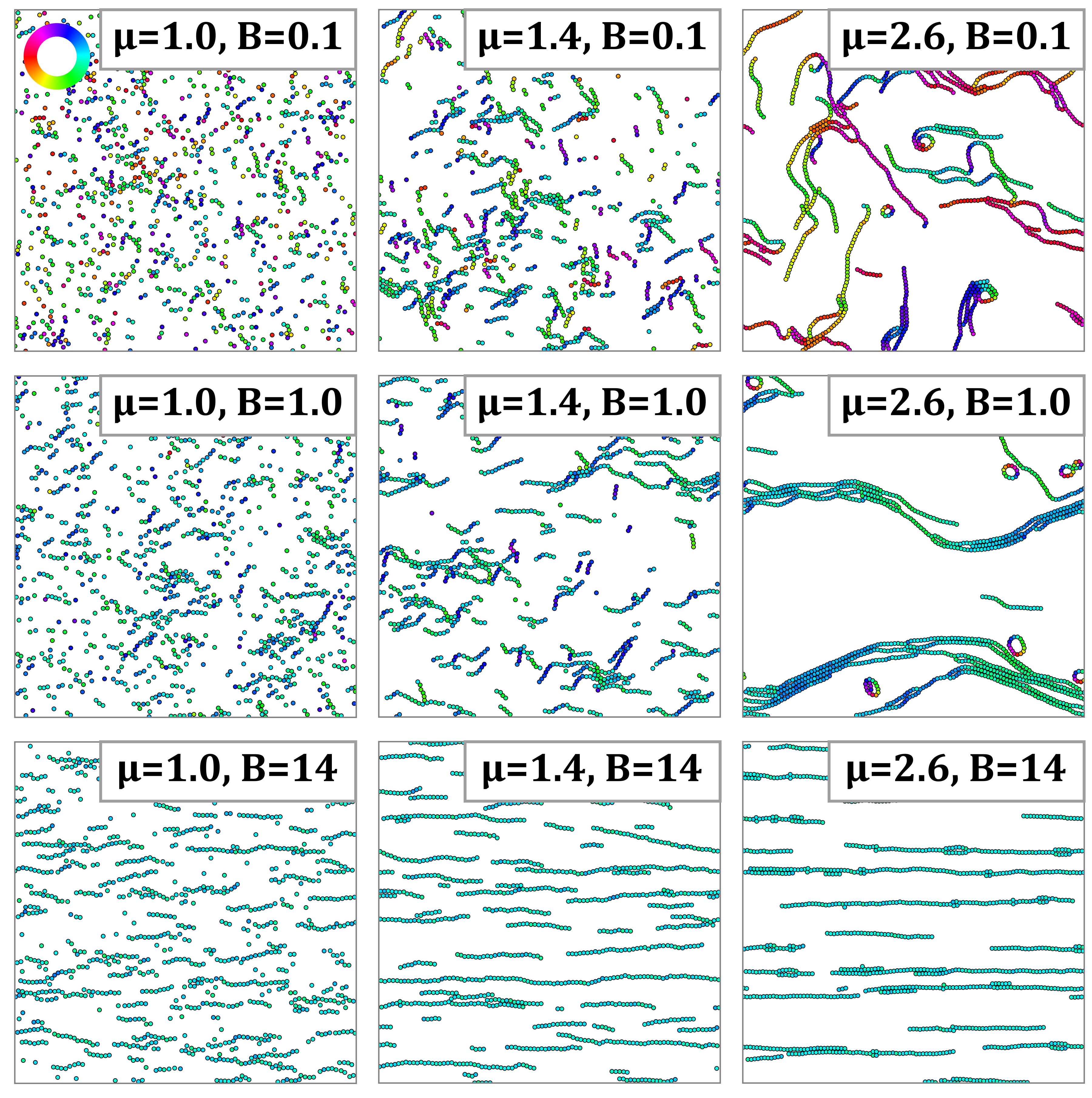}
    \caption{Collection of snapshots of the system at low density ($\Phi =0.13$) for intermediate self-propulsion speed $v_0=23$ for different magnetic field strengths $B$ (rows) and values for magnetic interaction strengths $\mu$ (columns). External magnetic field points from left to right. All snapshots have been taken at time $t=20$.}
    \label{fig:externalFieldOverviewLow}
\end{figure}

\Cref{fig:externalFieldOverviewLow} shows typical system configurations for three different values of the magnetic moments $\mu=$ 
\numlist{1.0;1.4;2.6} (columns from left to right) at these three field strengths ($B=$ \numlist{0.1; 1.0; 14} in the top, middle, bottom row, respectively) for particles with moderate self-propulsion speed ($v_0=23$).  The orientation of the particles is indicated by their color according to the color wheel. The magnetic field points from left to right, so alignment with the field can be seen by the cyan color of the particles.
Systems with a weak external magnetic field (top row) display similar configurations to what was observed in systems without external 
magnetic field (compare with \cref{fig:diagramOfStatesNoB}).
In systems with intermediate external magnetic fields (middle row), particles 
align with the direction of the external field (as seen by their cyan color).
In addition, particles assemble into chains 
for strong external magnetic fields ($B=14$). These chains often have defects such as branches or small loops.
Here, 
the average chain length is seen to increase with the magnetic moment of the particles, spanning the whole system 
size for strong magnetic interactions and strong external field strengths (for $\mu=2.6, \,B=14$, bottom right). These columnar 
ordered clusters 
are qualitatively similar to structures formed by non-equilibrium ferrofluids with external magnetic fields \cite{Flores1999}.  
In contrast to systems without an external magnetic field, these columnar clusters mostly stay evenly separated and do not 
combine into connected networks of chains or into broad bands (at the low density simulated here, broad bands are seen for higher density, as discussed below). 
In fact, we observed in systems with strong magnetic 
interactions ($\mu=2.6$) that an increase of the external magnetic field strength causes broad bands to separate into thinner 
bands with fewer lanes or into individual chains of dipolar particles (compare the cases for $\mu = 2.6$, $B=1.0$ and $\mu = 2.6$, $B=14$ in \cref{fig:externalFieldOverviewLow}). 

\begin{figure}[!tpb]
    \centering
    \includegraphics[width=\columnwidth]{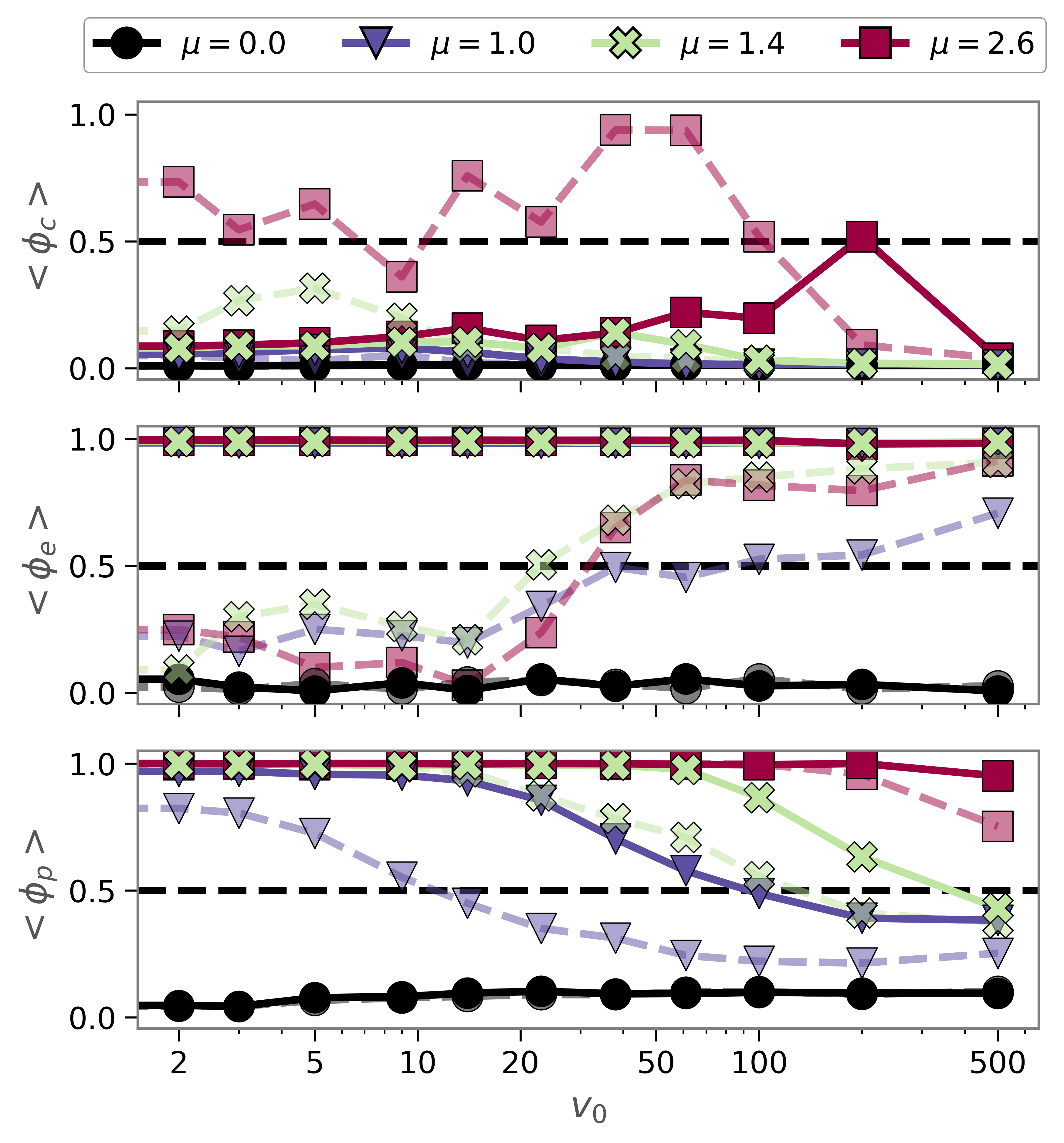}
    \caption{Three order parameters: fraction of largest cluster (top), polarization (center) and polymerization (bottom) against self-propulsion speed $v_0$ for four magnetic moments $\mu$ (colors). Solid color and solid lines show data for strong external magnetic field $B=14$, transparent colors and dashed lines show data with weak magnetic field strength $B=0.1$.  System density was set to $\Phi=0.13$.
    Cut-off for cluster formation set at $\delta_c=1.28$.}
    \label{fig:orderParamBComp}
\end{figure}

We analyzed the numerical simulations quantitatively by calculating the three global order parameters introduced above. 
They are plotted in \cref{fig:orderParamBComp} for different $\mu$ and $v_0$ and for strong and weak magnetic field. 
Clustering (with $\left\langle\phi_c\right\rangle>0.5$) is largely absent at the low density used here and is only observed for the highest magnetic moment and a weak magnetic field (corresponding to the bands or networks of chains seen in the absence of a magnetic field). The strong magnetic field suppresses clustering also under these conditions (red lines). 

The degree of polymerization  $\left\langle\phi_p\right\rangle$ (bottom) shows a slight 
increase with increasing field strength. The degree of polarization $\left\langle\phi_e\right\rangle$ (middle), not surprisingly, is high for strong external fields and independent of self-propulsion speed. By contrast, it shows a transition from no polarization to strong polarization at moderate propulsion speeds for the case of the weak magnetic field. $\left\langle\phi_p\right\rangle$ decreases with increasing 
self-propulsion speed $v_0$. 
 
\begin{figure}[!tpb]
    \centering
    \includegraphics[width=\columnwidth]{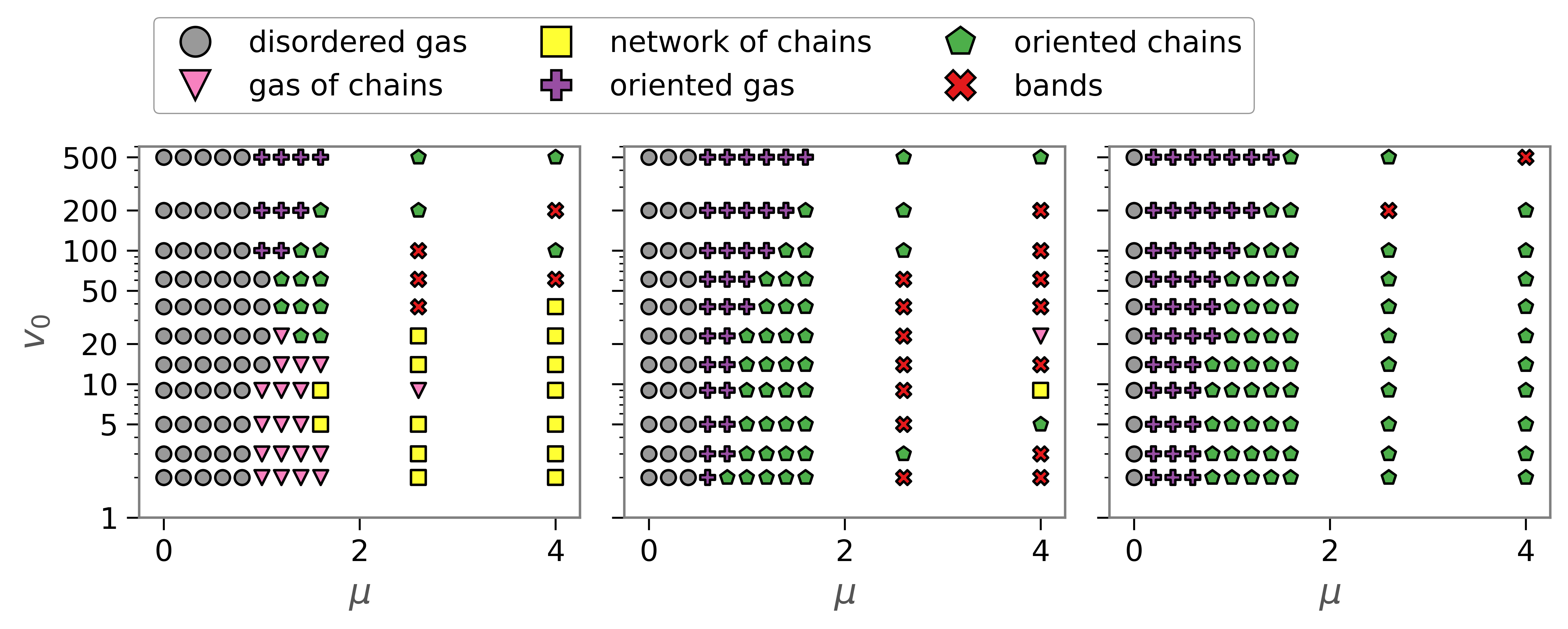}
    \caption{Diagrams of states for three different external magnetic field strengths $B=$ \numlist{0.1;1.0;14} (left to right). Classification of states is based on combinations of order parameter criteria (see \cref{tab:orderParamCombi}). System at low density ($\Phi =0.13$).}
    \label{fig:diagramOfStatesCompLowDensity}
\end{figure}

Using again the criteria based on the three order parameters, we classified the observed states of the system and summarized the results in a diagram of states, shown in
\cref{fig:diagramOfStatesCompLowDensity}.
The figure shows the diagrams of states for the three
different magnetic field 
strengths $B=$ \numlist{0.1;1.0;14} (left to right) and for the density $\Phi=0.13$. 
Based on these diagrams we can make three key observations: 
First, disordered gas states (gray circles) are strongly suppressed with increasing magnetic 
field strength. In addition, the onset of the oriented gas state (purple plus) shifts to lower values of $\mu$.
Second, intermediate as well as strong
magnetic fields (center and right diagram) suppress the formation of networks of chains (yellow square)
and instead promote the formation of oriented states, specifically the oriented gas (purple plus), oriented chains (green pentagon) and bands (red cross, for intermediate field strength).
Third, and most importantly, strong external magnetic fields suppress the formation of bands. For strong external magnetic fields 
(right diagram), two states dominate the diagram, the oriented gas (purple plus) and oriented chains (green pentagon). The line 
separating these two states resembles the corresponding line separating the gas and chain state in small confined systems of dipolar active particles investigated in previous work \cite{Telezki2020}. 

\begin{figure}[tb]
    \centering
    \includegraphics[width=.9\columnwidth]{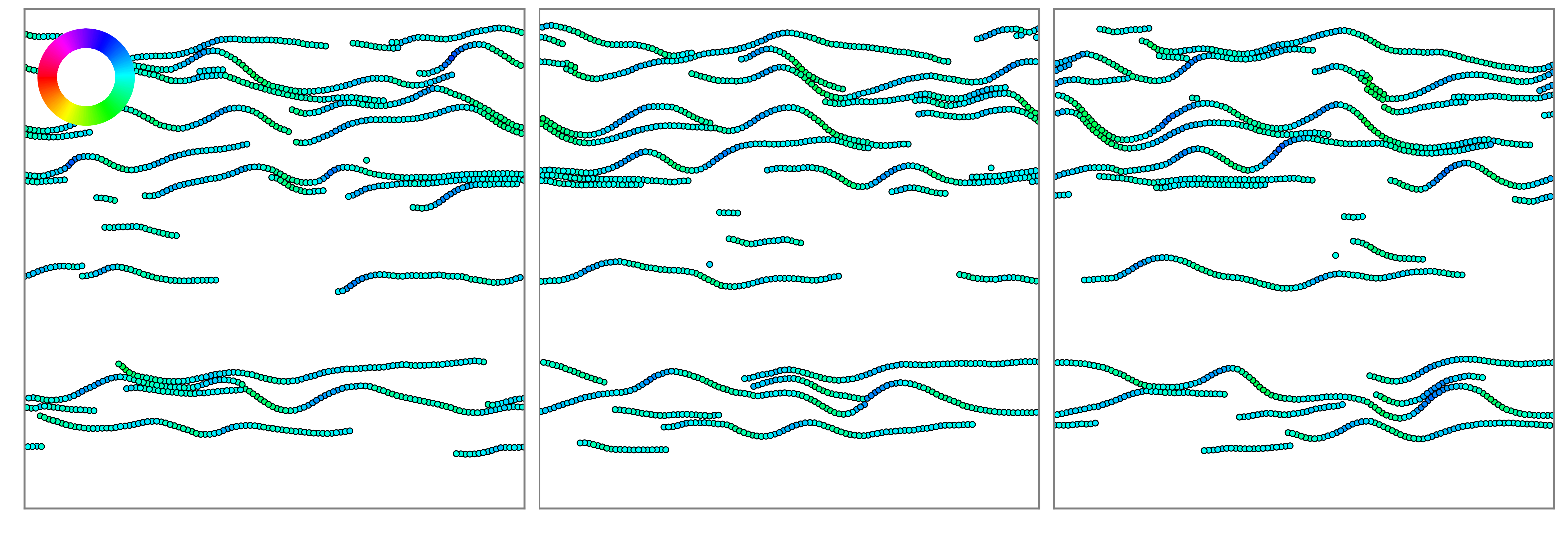}
    \caption{Series of snapshots showing oscillating chains for dipolar active particles with strong magnetic interactions $\mu=4$ and high self-propulsion speed $v_0=500$. Snapshots taken at times \numlist{26.46;26.56;26.66} (left to right), color wheel indicates orientations. External magnetic field points from left to right, magnetic field strength was set to $B=14$.}
    \label{fig:oscillations}
\end{figure}

For strong magnetic fields, band patterns were largely absent (right panel in \cref{fig:diagramOfStatesCompLowDensity}). We only found bands for two parameter combinations  
($\mu=2.6$, $v_0=200$ and $\mu$=4, $v_0=500$, indicated by the red crosses in \cref{fig:diagramOfStatesCompLowDensity}). In these cases we observed  the following noteworthy behavior, shown by the snapshots in \cref{fig:oscillations}.
Dipolar active particles that have self-assembled into long chains transiently show oscillations. The amplitude of these oscillations was seen to grow  towards their tail of a chain. We suspect
that such oscillations might be caused by a buckling instability:  In the chains, particles with strong dipolar interactions are packed rather
closely. The 
combination of high swimming speeds and translational noise can then cause a temporary compression and stress along the chain that
is  
resolved by buckling of the chain. When the dipolar interactions were reduced, long chains were seen to break into smaller segments. These 
breaks typically occurred near the tail of the chain, where the oscillations reach their maximum amplitude. This observation suggests 
that fragmentation induced by buckling may provide a mechanism for the coexistence of chains of different lengths. We note that the breaking of chains resembles the fragmentation of passive chains observed in magnetorheological suspensions driven by  a rotating field 
\cite{Melle2003,Melle2003a}.
In our simulations these oscillating chains are sees as a transient state towards band formation. At some point, individual 
oscillating chains meet. Then, alignment with the neighboring particles suppresses oscillations and a band starts to form. Again, similar lateral aggregation of individual chains into bands and columnar lanes has been observed in magnetorheological fluids \cite{Liu1995,Flores1999,Ivey2001,DeVicente2011}. 

\subsection{Effect of density}
\begin{figure}[!tpb]
    \centering
    \includegraphics[width=\columnwidth]{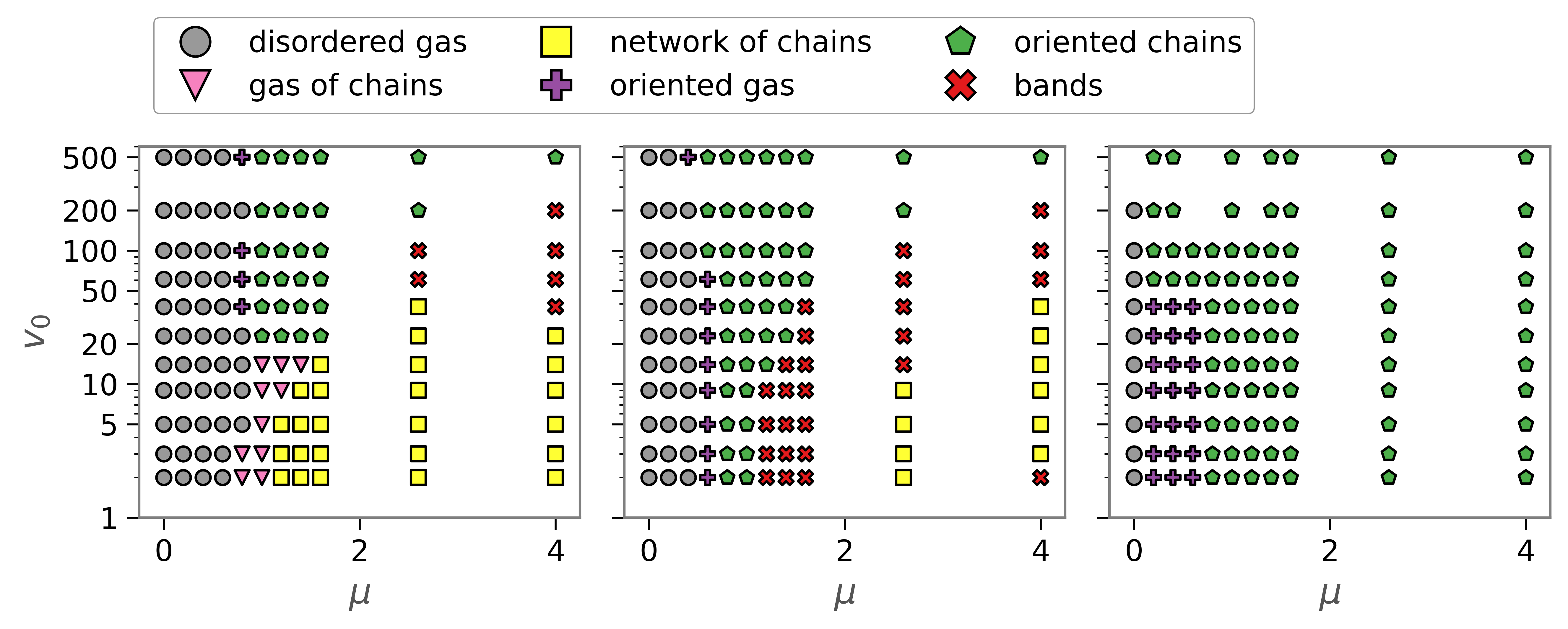}
    \caption{Diagrams of states for three different external magnetic field strengths $B=$ \numlist{0.1;1.0;14} (left to right). Classification of states is based on combinations of order parameter criteria (see \cref{tab:orderParamCombi}). System at intermediate density ($\Phi =0.23$).
    Cut-off for cluster formation set at $\delta_c=1.21$.}
    \label{fig:diagramOfStatesCompInterDensity}
\end{figure}

\begin{figure}[tpb]
    \centering
    \includegraphics[width=\columnwidth]{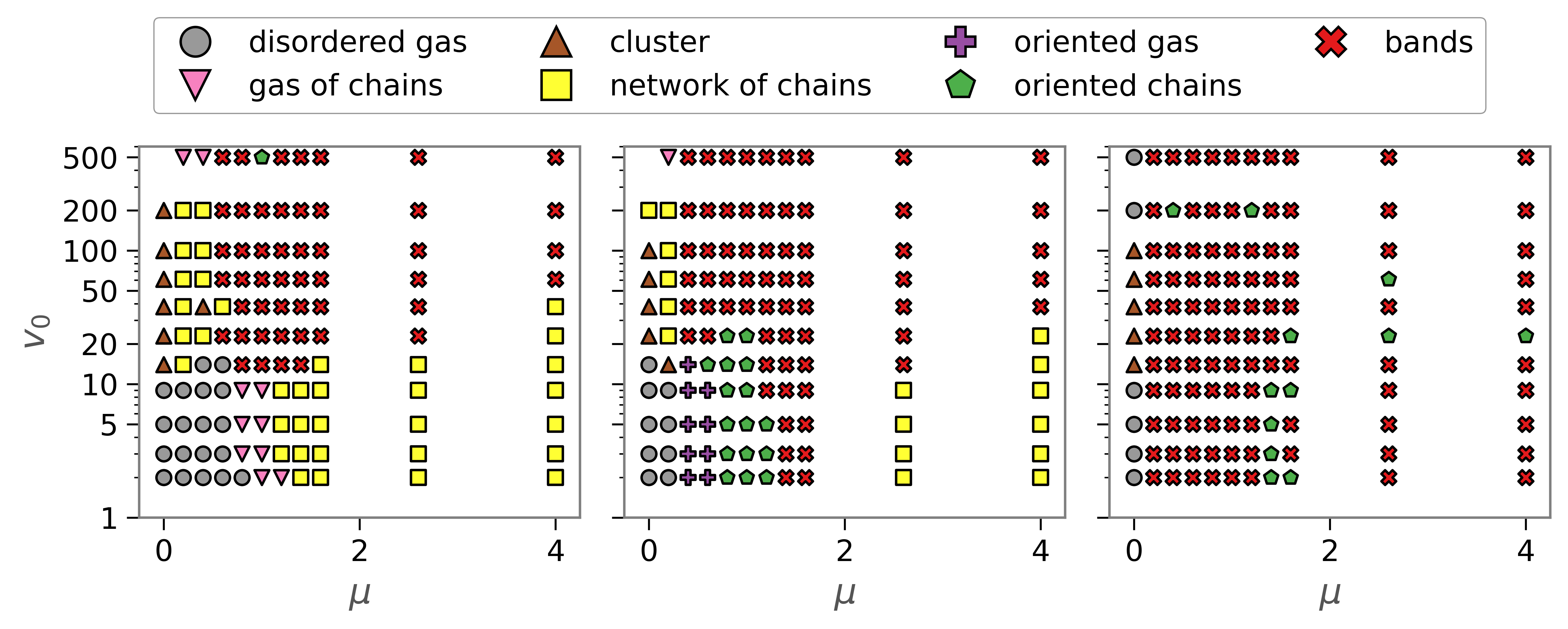}
    \caption{Diagrams of states for three different external magnetic field strengths $B=$ \numlist{0.1;1.0;14} (left to right). Classification of states is based on combinations of order parameter criteria (see \cref{tab:orderParamCombi}). System at high density ($\Phi =0.57$).
    Cut-off for cluster formation set at $\delta_c=1.05$.}
    \label{fig:diagramOfStatesCompHighDensity}
\end{figure}

To investigate the effect of the density of the system on the patterns formed by the active dipolar particles in a magnetic field, we simulated systems with two higher densities ($\Phi=0.23$ and $\Phi=0.57$) and repeated the calculation of the order parameters. The resulting state diagrams are shown in \cref{fig:diagramOfStatesCompInterDensity} (for $\Phi=0.23$) and in \cref{fig:diagramOfStatesCompHighDensity} (for $\Phi=0.57$). 

In systems with the intermediate density $\Phi=0.23$, we observed that intermediate external magnetic field strengths of $B=1$ do not suffice to suppress formation of complex networks  (yellow squares, compare the middle panel in 
(\cref{fig:diagramOfStatesCompInterDensity} with that for low density in \cref{fig:diagramOfStatesCompLowDensity}).
Because of the increase in density and therefore a reduced mean particle distance, short chains may connect and form a complex network.

Furthermore, the diagram of states indicates that strong external magnetic fields diminish the effect activity has in systems at 
intermediate density (\cref{fig:diagramOfStatesCompInterDensity}, right). Here, the line separating the oriented gas state 
(purple plus) and the oriented chains state (green pentagon) is independent of $v_0$. This observation is markedly  different from the 
previous observations for  systems without external magnetic fields and to systems at low density with external magnetic fields, where  activity had a strong influence on the collective behavior and structure formation. 

For systems of high density (our case with $\Phi=0.57$), three phenomena compete with one another: polar order, chain formation and MIPS, as pointed out by Liao et al.~\cite{Liao2020}. The
addition of an external magnetic field introduces global polar order and suppresses formation of complex networks of chains. Instead, we observed that strong external magnetic field strengths 
of $B=14$ clearly organize the system with high magnetic moments into columnar structures (see 
\cref{fig:externalFieldOverviewLow} $\mu=2.6, B=14$). Because of the 
increase in density, these columnar structures are made up of multiple lanes in off-register alignment and not of single lanes, 
as seen in systems at low density. 

In the diagram of states, we observed that the increased density causes the previously suppressed band states (red crosses) to 
reappear. In fact, bands dominate the corresponding 
diagram of states for strong external magnetic fields and are independent of activity (\cref{fig:diagramOfStatesCompHighDensity}, 
right). Off-register band formation has been previously observed in passive dipolar systems with constant external magnetic fields
\cite{Mohebi1996,Weis2003,Weis2005}.
In systems at high density, a clear discrimination between bands and columnar states can no longer be made. A limiting factor 
might be the finite system size, as observed in other systems of active particles \cite{Chate2008}.
\FloatBarrier
 
\subsection{Columnar clusters}
\begin{figure}[!tpb]
    \centering
    \includegraphics[width=.65\columnwidth]{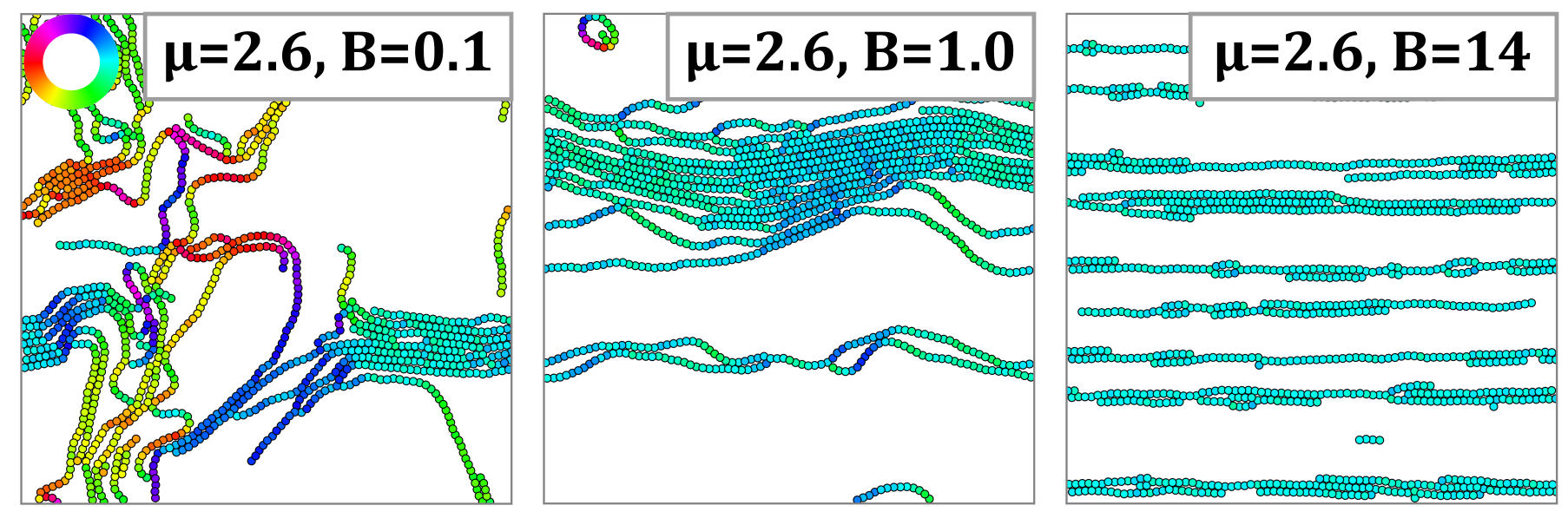}
    \caption{Snapshots of system with intermediate density and strong magnetic interactions show columnar clusters, parallel bands spanning the whole system in the direction of the field that shrink with increasing external field. The density is $\Phi=0.23$, the velocity $v_0=23$. }
    \label{fig:externalFieldStructures}
\end{figure}

At intermediate and high density and strong fields, the behavior of the system in dominated by oriented chains and bands. Snapshots of the system show additional structure in the scenarios classified as bands or oriented chains:  \cref{fig:externalFieldStructures} shows that as the magnetic field strength increases, a 
disordered network of chains first organizes into a broad 
band, which then, as the field strength is increased further, splits into parallel clusters of oriented chains. 
These structures are aligned in the direction of the external field and, for sufficiently strong 
magnetic interactions,  span the whole system. This effect seems to be more pronounced at the intermediate density $\Phi=0.23$, on which 
we will focus in the following.
Snapshots of the simulations indicate (see \cref{fig:externalFieldStructures})  that the number of 
lanes, i.e., of parallel chains in each cluster can vary and depends on the field strength. The clusters appear to be quite regularly spaced in the direction perpendicular to the external 
magnetic field. All these properties are reminiscent of magnetic field-induced columnar clusters  in ferrofluids \cite{Liu1995,Flores1999,Ivey2001}.

\begin{figure}[tb]
    \centering
    \includegraphics[width=.65\columnwidth]{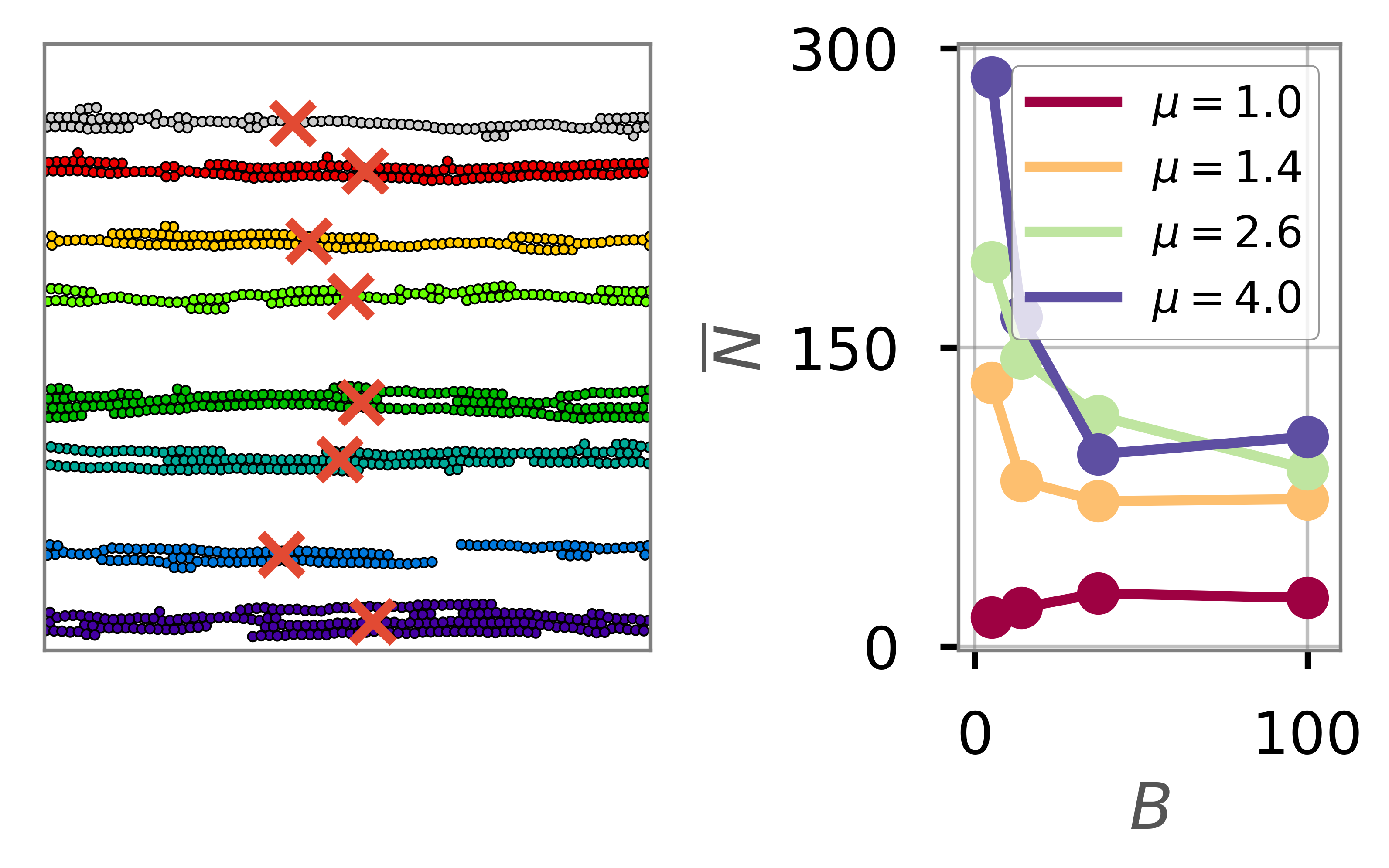}
    \caption{Left: snapshot of system at $B=14$, $\mu=2.6$, $v_0=14$ showing columnar clusters. Individual clusters are color coded, center of geometry is marked by a red cross. Right: average cluster size $\overline{N}$ against external magnetic field strength $B$ for particles with moderate self-propulsion speed $v_0=14$. System is at intermediate density $\Phi=0.23$. Colors indicate different magnetic moments of the dipolar particles. }
    \label{fig:clusterCOG}
\end{figure}

We characterized these columnar clusters with two parameters: the average number of lanes in a cluster 
$\overline {n_{\mathrm l}}$ and the columnar spacing $\overline {\Delta r_y}$, which is the  average distance between neighboring
cluster centers perpendicular to the external magnetic field. 
These two parameters were calculated as follows for simulations showing oriented chains or band (i.e., for simulations with $\left\langle\phi_e\right\rangle > 0.5$ and $\left\langle\phi_p\right\rangle > 0.5$):
First, we identified individual clusters (color coded in \cref{fig:clusterCOG}, left), using a density based cluster algorithm
(see Methods).
We then calculated the center of geometry (red crosses) and the length of each cluster in direction of the external magnetic field 
$l_x^i$. Averaging the distances between neighboring centers of geometry perpendicular to the direction of the external magnetic 
field while accounting for periodic boundary conditions, gives us the columnar spacing $\overline{\Delta r_{\mathrm {y}}}$. 
To approximate the number of lanes in each cluster, we divided the number of particles per cluster $n_i$ by the cluster extension in direction of the external magnetic field
$n_{\mathrm{l}}^i =\frac{n_i}{l_x^i}$. 

Indeed, the average number of particles per 
cluster $\overline{N}$ (\cref{fig:clusterCOG}, right) decreases rapidly with increasing magnetic field strength, confirming that clusters separate into smaller sized clusters as the magnetic field strength increases in systems 
with $\mu>1$. The average cluster size appears to converge to a fixed value for very strong external magnetic fields. 
\begin{figure}[tpb]
    \centering
    \includegraphics[width=.9\columnwidth]{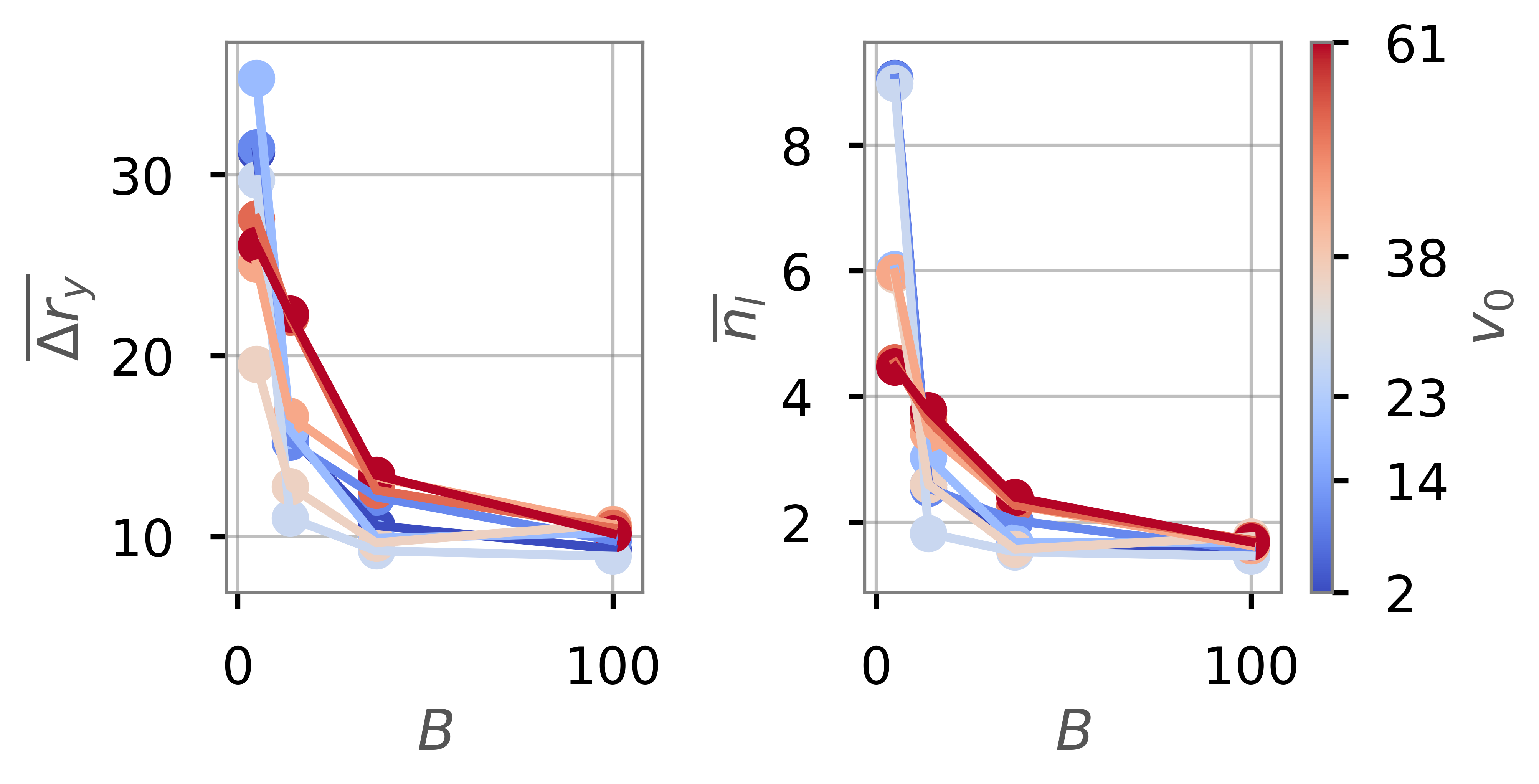}
    \caption{Columnar spacing $\overline{\Delta r_y}$ (left) and average number of lanes per cluster $\overline{n_{\mathrm l}}$ (right) against external magnetic field strength $B$. The magnetic moment was set to $\mu=4$ and density to $\Phi=0.23$. Colors indicate increasing self-propulsion speed of the particles. Data sampled over trajectories at times $t=18-20$ every $\Delta t=0.1$. Data points show systems that fulfill criteria $\left\langle\phi_e\right\rangle > 0.5$ and $\left\langle\phi_p\right\rangle > 0.5$.  }
    \label{fig:cogAndExternsionSmall}
\end{figure}
\Cref{fig:cogAndExternsionSmall} shows the columnar spacing $\overline {\Delta r_y}$ and the average number of lanes per column
$\overline {n_{\mathrm l}}$ for systems with strong magnetic interactions $\mu=4$. 
Both the 
columnar spacing and the number of lanes per cluster decay rapidly and approach a fixed value as the 
external magnetic field strength increases. Activity seems to have only a minor effect on this behavior. Columnar spacing has been reported to be independent of the external magnetic field in experiments with (passive)
ferrofluids \cite{Ivey2001}. Our result for active particles at for strong magnetic fields mirror this behavior.  
However, we observed a strong dependence of the columnar spacing in the regime of weak external magnetic fields. We note that in some cases, a weak external magnetic field seems to promote separation between columnar structures. 

As reported in \cite{Ivey2001}, columnar spacing is reduced with increasing density in ferrofluids. While we did qualitatively observe in movies of our 
simulations that the number of lanes in each column increases with density, more systematic investigation of density effects would be needed to make quantitative statements about the
influence of density on columnar structures.

\section{Conclusions}
In this study, we used extensive simulations to investigate the effect of a homogeneous external magnetic field on systems of active dipolar particles. Based on the combination of three order parameters, we determined diagrams of state for different magnetic field strengths to provide a comprehensive overview of the collective behaviors observed in such system. 
Consistent with previous work \cite{Liao2020,Telezki2020}, we see that dipolar interactions without an external field result in chains, which at higher densities form networks and bands. In our numerical experiments, motility-induced phase separation is only observed in the absence of dipolar interactions. 
Adding dipolar interactions, even with the weakest interaction strength we simulated, resulted in a network of chains instead. Therefore, we conclude that dipolar interactions likely prevent the directionally disordered jamming of particles and thereby inhibit the slow-down required for motility-induced phase separation. 

Under the influence of an external magnetic field oriented chains are dominant and bands are seen less frequently. For weak interactions, an oriented gas of individual dipoles is observed.  
Both states are only weakly dependent on the 
the level of activity and practically independent of it when a strong external magnetic field is present. 
This observation indicates that for strong field, the magnetic interactions dominate over activity, such that systems of active dipolar particles resemble passive ferrofluids more than classical (non-dipolar) active Brownian particles. This observation for strong field stands in stark contrast to the 
chain formation observed in systems without external magnetic fields, where the activity plays a 
crucial role in collective phenomena \cite{Liao2020,Telezki2020,Maloney2020}. 

Two observations may merit further research in future work: (i) We observed oscillating chains as a transient state towards band formation in our system. We suspect that 
these oscillations are caused by a buckling instability and that  fragmentation of long chains due to buckling allows chains of different lengths to coexist in dilute systems, despite  
high magnetic moments. The self-assembled active dipolar chains we study here have some similarities to active polymers, in which the 
monomers are linked by a spring potential \cite{winkler2020physics}. In that case, buckling has been studied both theoretically and experimentally \cite{Prathyusha2018,abbaspour2023effects,kurjahn2023quantifying,vliegenthart2020filamentous} and was shown to be controlled by the flexure number, which characterizes the relation between active propulsion and bending rigidity
\cite{Prathyusha2018,abbaspour2023effects}.

(ii) We observed that active dipolar particles form columnar clusters that resemble structures induced by external magnetic fields in (passive) ferrofluids \cite{Liu1995,Flores1999,Ivey2001}. A quantitative analysis showed that the 
number of lanes per column and the columnar spacing decreases rapidly with increasing magnetic field strength, but become constant for strong fields. The strong-field behavior is similar to observations in ferrofluids, where 
columnar spacing has been reported to be independent of the external magnetic field strength \cite{Liu1995,Ivey2001}.
Further features of the system and experimental protocols (such as the application of the external magnetic field, presence of hydrodynamic interactions, polydispersity and boundary conditions) may have an effect on this result. In a first step, we have shown here that activity does not strongly affect separation between columns. In future studies, the effect of density could be investigated in more detail and compared to observations for ferrofluids \cite{Liu1995}.

Our results together with those from previous studies \cite{Liao2020,sese2022impact} show that a preferred orientation (introduced by the external field) coupled to activity can significantly change the observed collective behaviors. In our model, activity is introduced as an 
active force acting in the direction of the magnetic moment of the particle. Therefore, orientation competes with the formation of structures based on random or conflicting directions of self-propulsion such as MIPS. These observations indicate that the details  of how activity is introduced into the 
system and whether or how the direction of self-propulsion is coupled to the direction of the dipoles should be important for the type of patterns and collective phenomena that are observed. They suggest that even richer collective phenomena are possible if there is loose coupling between the dipolar and the self-propulsion direction of the active particles.

\section*{Data and code availability}

Data and code are available via GRO.data \cite{data,code}. 

\begin{acknowledgments}
This work was supported by the Deutsche Forschungsgemeinschaft (DFG, German Research Foundation – project IDs 253375392; 446142122). 
The simulations were run on the GoeGrid cluster at the University of G\"ottingen, which is supported by DFG (project IDs 436382789; 493420525) and MWK Niedersachsen (grant no.\ 45-10-19-F-02).
\end{acknowledgments}


\bibliography{apssamp}

\end{document}